\documentstyle[12pt,aasms4,psfig]{article}

\lefthead{Gioia et al.}
\righthead{Evolution of the NEP cluster X-ray Luminosity}

\def\arcmin{\hbox{$^\prime$}}

\def\arcsec{\ifmmode^{\prime\prime}\;\else$^{\prime\prime}\;$\fi}

\begin{document}

\title{Cluster Evolution in the ROSAT North Ecliptic Pole Survey}

\author{I. M. Gioia\altaffilmark{1,2}, J. P. Henry\altaffilmark{2},
C. R. Mullis\altaffilmark{2}}

\affil{Institute for Astronomy, University of Hawaii, 2680 Woodlawn 
Drive, Honolulu, HI 96822}

\and

\author{W. Voges, U. G. Briel, H. B\"ohringer}

\affil{Max-Planck-Institut fur Extraterrestrische Physik,
Giessenbachstrasse Postfach 1603, Garching, D-85740, Germany}

\and
 
\author{J. P. Huchra}

\affil{Harvard-Smithsonian Center for Astrophysics, 60 Garden Street,
Cambridge, MA, 02138}

\altaffiltext{1}{Home Institution: Istituto di Radioastronomia del CNR,
Via Gobetti 101, I-40129, Bologna, ITALY}

\altaffiltext{2}{Visiting Astronomer at Canada-France-Hawaii Telescope,
operated by the National Research Council of Canada, le Centre
National de la Recherche Scientifique de France and the University
of Hawai$'$i, and at the W. M. Keck Observatory, jointly
operated by the  California Institute of Technology, the
University of California and the National Aeronautics and Space
Administration.}

\begin{abstract}

The deepest region of the ROSAT All$-$Sky Survey, at the North 
Ecliptic Pole, has been studied to produce a complete 
and unbiased X-ray selected sample of clusters of 
galaxies. This sample is used to investigate the nature of cluster 
evolution and explore potential implications for large-scale 
structure models. The survey is 99.6\% optically identified. 
Spectroscopic redshifts have been measured for all the 
extragalactic identifications. In this Letter, first results on cluster 
evolution are presented based on a comparison between the number of 
the observed clusters in the North Ecliptic Pole survey and the 
number of expected clusters assuming no-evolution  models. 
At $z>0.3$ there is a  deficit of  clusters with respect to the local
universe which is significant at $>$ 4.7$\sigma$. The evolution 
appears to commence at L$_{0.5-2.0} > 1.8\times10^{44}$ erg s$^{-1}$ 
in our data. The negative evolution goes in the
same direction as the original EMSS result, the results from the 
160 deg$^{2}$ survey by  Vikhlinin et al. (1998) and the recent 
results from the RDCS (Rosati et al.\ 2000). At lower redshifts there 
is no evidence for  evolution, a result in agreement with these and
other cluster surveys.

\end{abstract}


\section{Introduction}

Since the X-ray surveys of the early 1980s, there has been remarkable 
progress in the construction of 
homogeneous samples of X-ray  selected clusters over large redshift  
baselines. X-ray surveys are sensitive enough to detect objects at 
redshifts of order unity (e.g. MS1054$-$03 at z$=$0.83, \cite{gio94};
RXJ0848.9+4452 at $z=1.25$, \cite{ros99}).
Large optical telescopes have the capability to locate them
and measure their redshifts.
The properties of clusters in this high redshift range constrain 
cosmological parameters since some evolution is expected in the 
look-back time that approaches half the age of the universe.
Observations of the highest redshift systems provide the longest lever 
arm in the attempt to evaluate the evolution of the bulk properties 
of clusters. In 1986 Kaiser predicted that if one assumes a power-law  
fluctuation power spectrum and  gas heated only by adiabatic compression
during the collapse of the cluster  dark matter halo,
then the comoving number density of clusters 
should increase, for clusters with fixed X-ray luminosity, rather 
than decrease  with redshift. The observed lack of high-z 
and X-ray luminous clusters reported in the early years
(\cite{gio90a}; \cite{edg90}; \cite{hen92}) is in conflict 
with this model. However, when models with additional  heat (or 
entropy) are assumed then negative evolution is predicted for only the 
very bright clusters (L$_{X} \geq 5 \times 10^{44}$ erg s$^{-1}$), 
with  little evolution at lower luminosities (\cite{kai91}; 
\cite{evr90}; \cite{evhen91}; among others).

What do we know today about the evolution of the cluster X-ray
luminosity function (XLF)?
All the determinations of the cluster XLF derived from existing
surveys are in agreement for low redshifts ($z<0.3$) and for low 
luminosities (L$_{0.3-3.5}<5\times10^{44}$ erg s$^{-1}$) at high
redshift. However, there 
is no unanimity regarding the most luminous and most distant X-ray  
clusters known, though there may be an accumulation of evidence in
favor of evolution. The Extended Medium Sensitivity Survey (EMSS; 
\cite{gio90b}; \cite{sto91}; \cite{mac94}; \cite{gio94}) was the first 
dataset where negative evolution was detected  (3$\sigma$ confidence 
level) between low redshift (0.14$< z <$0.2) and high redshift 
(0.3$< z <$0.6) clusters, but only at luminosities 
L$_{0.3-3.5} > 5\times10^{44}$ erg s$^{-1}$ (\cite{gio90a}; \cite{hen92}).
Nichol and collaborators (1997) questioned these results.
They revisited the EMSS XLF using a combination 
of the original Einstein Observatory data and  ROSAT follow-up 
observations and reduced the significance of evolution for $z<0.5$ to  
only the 1$\sigma$ level.  However one of the latest studies of
evolution of the EMSS sample by the same group, using the same 
data (\cite{rei99}), finds a deficit of a factor of 4$-$5 for 
X-ray clusters at redshift  $0.4<z<0.9$  at luminosities 
above $7\times 10^{44}$ erg s$^{-1}$ in 0.3$-$3.5 keV band.

The controversial  issue of the cluster XLF evolution inspired many 
EMSS-style cluster surveys, all based on ROSAT archival deep pointing 
images.  Each one of these surveys covers an area of sky of less than 
200 deg$^{2}$, significantly  less than the $\sim800$ deg$^{2}$ 
of the original EMSS (see Henry et al. 2001 for details on the ROSAT
surveys sky coverages), but with sensitivities almost an order 
of magnitude deeper
at the faint end ($\sim 1.8\times10^{-14}$ erg cm$^{-2}$ s$^{-1}$ vs 
$\sim 1.3\times10^{-13}$ erg cm$^{-2}$ s$^{-1}$ in 0.3$-$3.5 keV
\footnote{The conversion from 0.5$-$2.0 keV to 0.3$-$3.5 keV is 
a multiplicative factor of 1.8, assuming a Raymond-Smith model 
with a  kT$=$6.0 keV and the standard 0.3 solar abundance.}). 
Results are available for most of the ROSAT surveys and are summarized 
here. Collins et al.  (1997) and Burke et al. (1997) find no  
evolution in the Southern  SHARC survey (Serendipitous High-Redshift 
Archival Rosat Cluster) based on 16 clusters in the redshift range 
0.3$-$0.7 and luminosities up to 3 $\times 10^{44}$ erg sec$^{-1}$ in 
the 0.5$-$2.0 keV band. However, in the most recent Bright SHARC sample,
(which is a different sample, see \cite{rom00} for details)
Nichol et al. (1999) find a deficit of high z
clusters compared to what is expected from a non-evolving XLF. 
No significant evolution is found in the RDCS sample (Rosat Deep
Cluster Survey; Rosati et al. 1995 and 1998) for
L$_{0.5-2.0}<10^{44}$ erg s$^{-1}$ out to z$\sim$0.8. 
By adding the most distant clusters out to $z\simeq1.2$
Rosati et al. (2000) present evidence for negative evolution
of the XLF. The 160 deg$^{2}$  survey by Vikhlinin et al. (1998), 
detects a deficit of $z>0.3$ clusters with 
L$_{0.5-2.0} >3\times10^{44}$ erg s$^{-1}$ of a factor $3-4$ (see also
\cite{vik00}). On the other hand, Jones et al. (1998; 2000)
exclude a strong negative evolution of the most luminous and distant 
clusters extracted from the WARPS  (Wide Angle Pointed Rosat Survey; 
\cite{sch97}). Thus, except for the WARPS results, a consistent 
picture emerges from the existing X-ray surveys. No significant 
evolution is found for the low luminosity clusters up to $z\sim0.8$, 
but evolution of the most luminous clusters to $z > 0.8$
is not excluded.

In this Letter we present evidence regarding the evolution  of 
the cluster population based on the ROSAT All-Sky Survey 
observations  around the North Ecliptic Pole region (NEP). By 
comparing the number of detected NEP clusters with predictions 
from no-evolution models, we find  agreement between the results 
reported here and the original findings of the EMSS. Throughout this 
Letter we use H$_{0}=$50 km s$^{-1}$ Mpc $^{-1}$ and q$_{0}=$0.5.

\section{The ROSAT NEP survey}

The ROSAT NEP survey covers a $9^{\circ} \times 9^{\circ}$ region of the 
deepest area of the ROSAT All-Sky Survey (RASS; \cite{tru91}; 
\cite{vog99}) where the scan circles converge and the effective 
exposure  time approaches  $40$ ks (note that only 50\% of the sky in the 
RASS has an exposure time $>$ 400 s). An overview  of the ROSAT NEP survey 
and of the X-ray data can be found in Henry et al. 2001 and Voges et 
al. 2001.
The main difference between the NEP survey and the existing X-ray 
serendipitous  cluster surveys described above is that the NEP survey is 
both deep (median flux limit is f$_{0.5-2.0}=7.8\times10^{-14}$ erg
cm$^{-2}$ s$^{-1}$) and 
also covers a contiguous area of sky. Thus our database can be used to
examine large-scale structure in the cluster distribution.
A concentration of 21 groups and clusters was indeed found during the 
analysis of the NEP sources. The discovery of this supercluster is
reported in another Letter in this issue (Mullis et al. 2001).

Here we briefly mention the main properties of the NEP survey.
A total of 445 X-ray sources were detected with flux
determinations  $>4\sigma$ in the 
$0.1-2.4$ keV band using the RASS-II processing (described in detail
in \cite{vog99}). We have spectroscopically identified all but two
sources in the survey. Redshifts  have been measured for the 
extragalactic population (except for the very few previously cataloged
objects). We have extracted a complete and 
unbiased  sample of 64 galaxy clusters. Nineteen clusters have 
a redshift greater than 0.3 with the highest at z$=$0.81.

\section{The ROSAT NEP Cluster XLF}

\begin{figure}
\centerline
{\vbox{
\psfig{figure=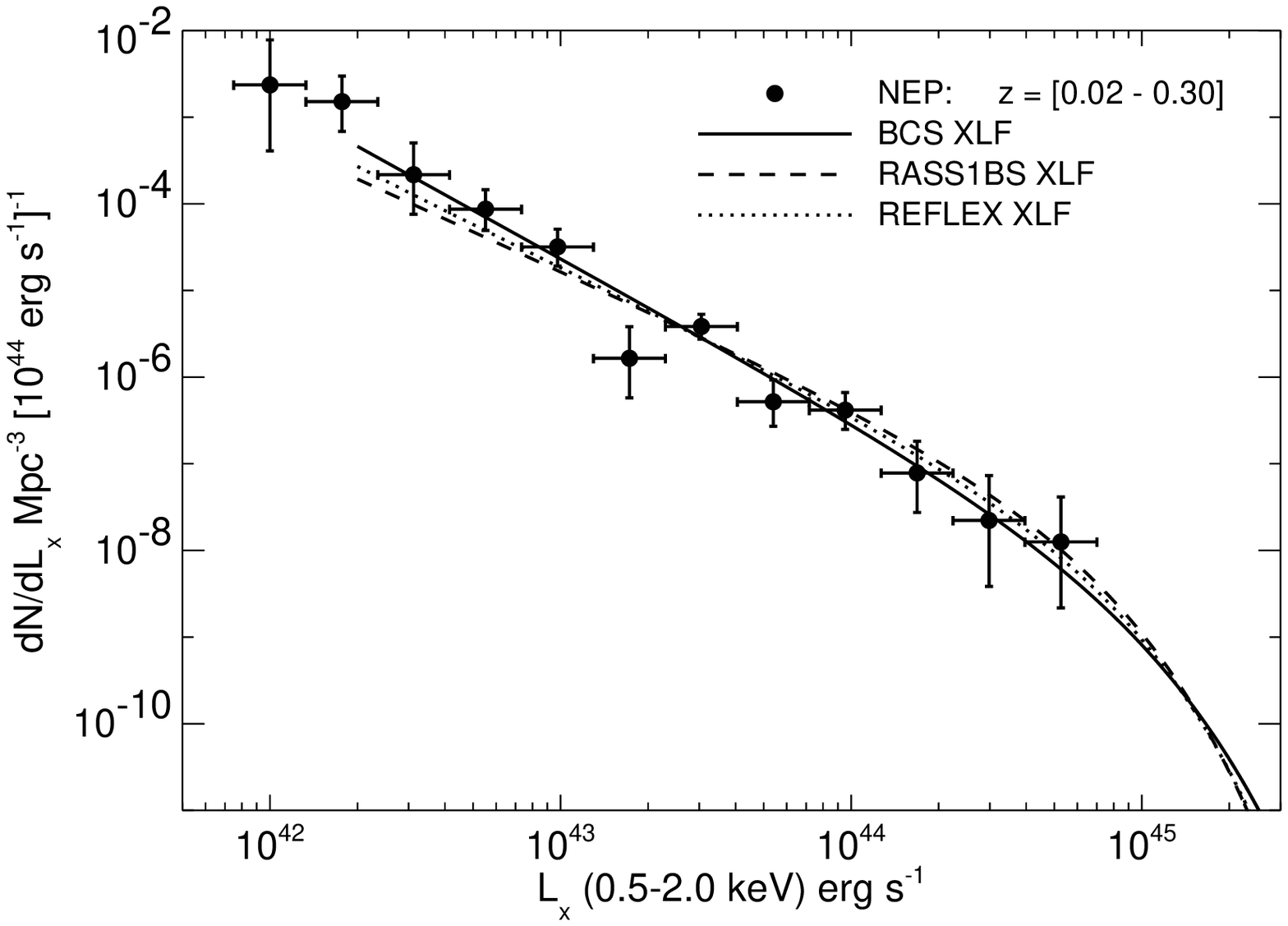,height=8.5cm,width=10.cm,angle=0.}
\psfig{figure=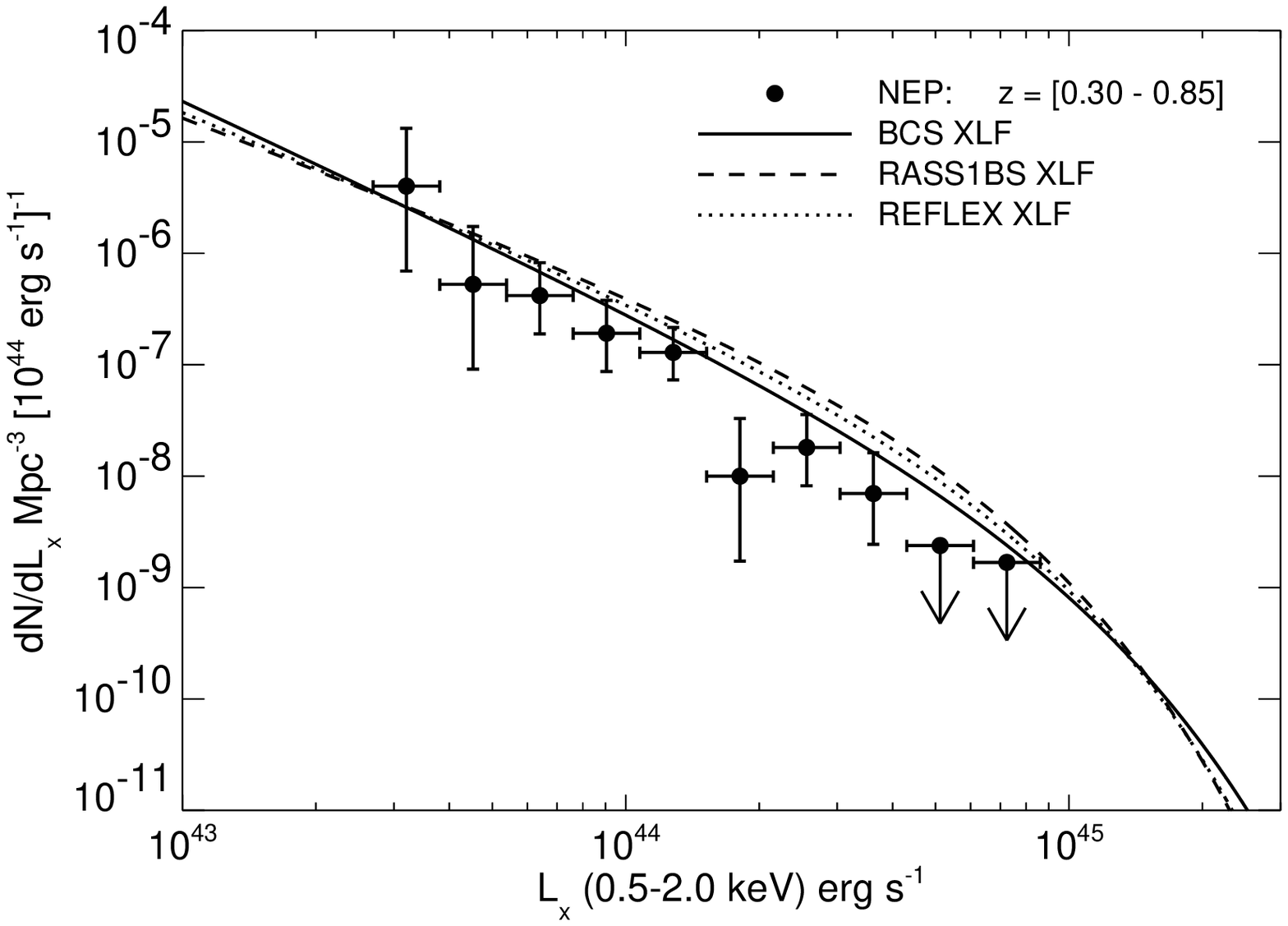,height=8.5cm,width=10.cm,angle=0.}
}}
\caption{The X-ray luminosity function for the NEP clusters in the range 
$0.02<z<0.3$ (top panel) and for clusters in the range $0.3<z<0.85$ (bottom
panel). The local XLF curves of the RASS1 (dashed), REFLEX (dotted) and BCS 
(solid) are overplotted.
\label{xlf_lo}
}
\end{figure}

The local cluster XLF has been derived by several authors (\cite{bur96};
\cite{ebe97}; \cite{deg99} among others) since it provides a crucial 
reference for cluster evolutionary studies at high redshift.  
Two determinations of the NEP cluster XLF have been computed, 
one for clusters with z$<$0.3 and one for clusters with z$>$0.3. A 
non-parametric representation of the differential XLF has been obtained 
following the 1/V$_{a}$ method (\cite{ab80}). 
We use three local (z$<$0.3) luminosity functions derived from the southern 
hemisphere RASS1 Bright Sample (\cite{deg99}), the southern hemisphere
REFLEX sample (B\"oringer et al. 1998; Guzzo et al. 1999) and  the 
northern hemisphere BCS sample (\cite{ebe97}). All three samples
were selected from the ROSAT All-Sky Survey with somewhat
different selection procedures and data analysis techniques.
The RASS1, REFLEX and BCS contain  126, 452, and 199 cluster
respectively. The top panel in Figure~\ref{xlf_lo} shows the NEP cluster 
luminosity function in the redshift
range (0.02$<$z$<$0.3) with the local XLFs overplotted. 
The vertical error bars are derived from Poisson statistics on the 
number of clusters in each bin while the horizontal 
error bars represent the logarithmic bin width. In the bottom panel 
of Figure~\ref{xlf_lo} the distant cluster (0.3$<$z$<$0.85) XLF is 
shown.  While the NEP local XLF is in agreement with 
independent determinations by other authors using different datasets,
the distant cluster XLF shows deviation from both the RASS1,
the REFLEX and BCS XLFs at luminosities greater than 
$>1.8\times10^{44}$  erg s$^{-1}$, the center of our first luminosity
bin significantly below the low redshift XLF.

\section{A deficit of clusters at high z and high L$_{X}$}

The number of observed clusters has been compared to the number of
expected clusters, assuming no-evolution models.
We have used a 5 \arcmin ~radius detect cell to derive background 
subtracted counts in the  ROSAT $0.1-2.4$ keV band. A King profile 
with $\beta$=2/3 and a core radius of 0.25 Mpc convolved with 
the RASS PSF has been integrated out to infinity to compute total 
cluster fluxes, quoted in the hard $0.5-2.0$ keV band.
K-corrected luminosities are computed assuming a temperature 
based on the L$_{X}-T_{X}$ relation of White et al. (1997). 
The three local luminosity functions derived from the RASS1, 
REFLEX and BCS have been folded through the NEP sky coverage and then
integrated in the appropriate redshift and luminosity ranges.
The ranges of integration were $z<0.3$ and $0.3<z<0.85$ in redshift,
and $3\times10^{42}-10^{47}$ erg s$^{-1}$ in luminosity (0.5-2.0 keV).
For the $z<0.3$ redshift range, the number of clusters
expected from the three local samples and observed in the NEP are
consistent, with the the significance of difference equal to 
$0.1-0.2\sigma$. For the $0.3<z<0.85$ range a value of 65.5 
clusters is expected according to the RASS1, a value of 55.9 according 
to the REFLEX and a value of 44.2 according to the BCS. Only 19 NEP 
clusters are observed in the same redshift and luminosity ranges. 
The significance of deviation is 6.4$\sigma$, 
7.2$\sigma$ or 4.7$\sigma$
depending on which of the three local XLF sample determinations is 
considered (see Table 1 for details). This result goes in the 
same  direction as the evolution derived from the EMSS survey. 

\section{The NEP Cluster LogN($>$S)-logS relation}

\begin{figure}
\epsscale{1.}
\plotone{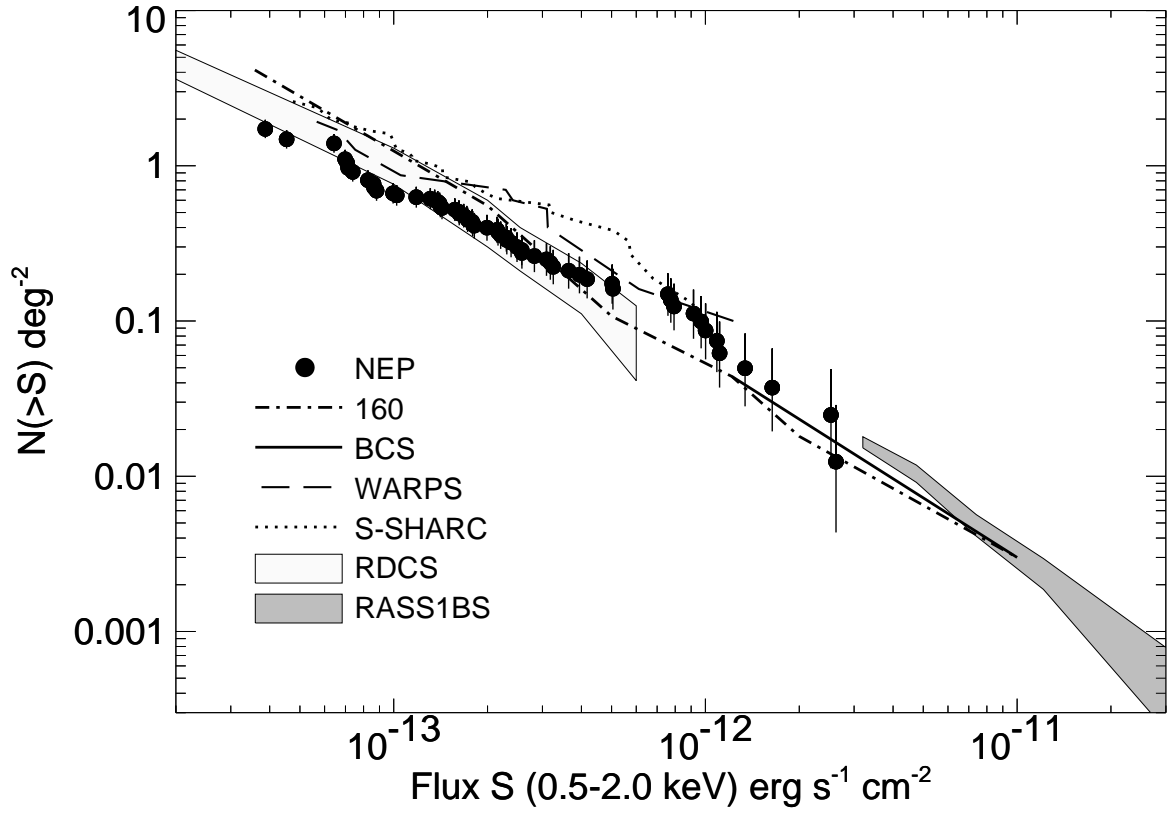}
\caption{The observed cumulative number counts from the NEP cluster survey
(solid points).  The logN($>$S)-logS for other surveys are also shown.
\label{lnls}
}
\end{figure}

The cumulative (integral) number counts of galaxy clusters are a less 
stringent constraint than the differential cluster XLF as a test of
evolution of the population. It is shown to verify
reliability of sky coverage plus completeness of identifications.
It is useful to compare the number counts of the NEP clusters 
with the existing  LogN($>$S)-logS relations derived by other
investigators. The observed cumulative  LogN($>$S)-logS for the NEP 
clusters is given in  Figure~\ref{lnls}.  Shown are also the 
number counts derived  from the 160 deg$^{2}$ (Vikhlinin et al. 1998),
the BCS (Ebeling et al. 1997), the WARPS (Jones et al. 1998), the S-SHARC 
(Burke et al. 2001), the RDCS (Rosati et al 1998), and the RASS1-BS (De
Grandi et al. 1999). The NEP cluster number counts 
are in agreement  within the errors with all the other independent 
determinations. 

\section{Conclusions}

In the last nine years we have spectroscopically identified all but
two of the 445 NEP sources, thus reaching an identification rate of 
99.6\%.  The resulting complete cluster sample contains 64 clusters, 
19 at a redshift $>$ 0.3. We find evidence for a deficit 
of clusters at  L$_{0.5-2.0}>1.8\times10^{44}$ erg s$^{-1}$ and 
$z>0.3$ compared to expectations from a non-evolving XLF. These
results go in the same direction as those of other surveys. The EMSS, 
160 deg$^{2}$, SHARC, RDCS and NEP surveys are now reporting
negative evolution at varying levels of significance from 
$\sim$ 1$\sigma$ to greater than 5$\sigma$. 
The NEP survey (this Letter) excludes the no-evolution model at 
at  4.7$\sigma$, 6.4$\sigma$ or 7.2$\sigma$, depending on the local 
XLF considered; the 160 deg$^{2}$ survey (Vikhlinin et al.\ 2000)
claims a deficit with respect to the BCS significant at the
3.5$\sigma$; the Bright SHARC survey (Nichol et al. 1999) claims a deficit
of clusters significant at 1.7 (1.1$\sigma$) with respect to the 
no-evolution prediction of the RASS1 (BCS); the RDCS (Rosati et al.\ 2000)
finds a departure of their best fit model from the no-evolution BCS 
prediction significant at more than 3$\sigma$; finally an analysis
of ASCA data of EMSS clusters (Henry 2001), using the same approach 
adopted here, finds a deficit of EMSS clusters compared to the RASS1 
and BCS predictions significant at $>$ 5$\sigma$. The only completely, 
or nearly  completely, analyzed cluster  sample that does not find 
negative evolution is the WARPS. Evidence is  thus accumulating in 
favor of evolution at the high luminosity end of the XLF at high z.
However, given the small number statistics and incomplete 
optical follow-ups of some of the existing  serendipitous cluster 
surveys, the issue  has not been completely resolved yet.
We stress that all but two NEP sources have been optically
identified. This practically complete identification rate
gives us confidence that the deficit of clusters seen is not due 
to the fact that clusters have been missed.
Larger and better characterized samples are needed to address with
greater confidence the evolutionary properties of the cluster
population. The large X-ray telescopes now in orbit (Chandra and 
XMM-Newton) will not be able to provide in the near future
serendipitous surveys of size comparable to what already exists.
However, with their high throughput 
and energy resolution it will be possible to obtain  a spectral 
determination for a very large number of clusters, and thus 
more reliable flux measurements which will help reducing some of the 
systematic uncertainties in the derived XLF and number counts of 
clusters of galaxies. 

\acknowledgments

We benefitted from helpful discussions and comments with A. Wolter, 
B. Tully and T. Maccacaro. We acknowledge the continuous support of 
the University of Hawai$'$i TAC and of the staff at the UH 2.2m, 
the CFH 3.6m and the Keck 10m telescopes.
Partial financial support for this work comes from NSF (AST91-19216
and AST95-00515), NASA (NGT5-50175, NASA-STScI GO-5402.01-93A and 
GO-05987.02-94A), NATO (CRG91-0415) and from the Italian Space Agency ASI.

\clearpage
\begin{deluxetable}{rcccc}
\tablenum{1}
\tablecaption{Predicted vs Observed Numbers of clusters \label{tbl-1}}
\tablehead{
\colhead{L$_{X}$ range} &  \colhead {Redshift} & \colhead {No-evolution} & \colhead{Observed} & \colhead  {Significance of} \\
\colhead{erg s$^{-1}$} &  \colhead{} & \colhead{Predictions} & \colhead{in NEP} & {difference ($\sigma$)} \\
\colhead{} &  \colhead{} & \colhead {RASS1\tablenotemark{a}} &  & {RASS1} \\
\colhead{} &  \colhead{} & \colhead {REFLEX\tablenotemark{b}}  & & {REFLEX} \\
\colhead{} &  \colhead{} & \colhead {BCS\tablenotemark{c}}      & & {BCS}
}
\startdata
$3\times10^{42} - 10^{47}$ &  z$<$0.3     & 44.5$\pm$4.0\tablenotemark{d} &  45$\pm$6.7\tablenotemark{e} & 0.1  \nl 
                           &              & 43.4$\pm$2.0 &                  & 0.2   \nl 
                           &              & 44.2$\pm$3.1 &                  & 0.1   \nl
$3\times10^{42} - 10^{47}$ & 0.3$<z<$0.85 & 65.5$\pm$5.8 &   19$\pm$4.4     & 6.4   \nl
                           &              & 55.9$\pm$2.6 &                  & 7.2   \nl
                           &              & 44.2$\pm$3.1 &                  & 4.7   \nl
\tablenotetext{a} {126 clusters}
\tablenotetext{b} {452 clusters}
\tablenotetext{c} {199 clusters}
\tablenotetext{d} {The fractional errors in this column are the inverse 
of the square root of the total number of objects in the respective samples}
\tablenotetext{e} { The errors in this column are the  square root of the
number of clusters observed}
\enddata
\end{deluxetable}

\clearpage

\clearpage

\end{document}